\documentstyle[12pt]{article}
\newlength{\extraspace}
\newlength{\extraspaces}
\newcommand{\be}{\begin{equation}
\addtolength{\abovedisplayskip}{\extraspaces}
\addtolength{\belowdisplayskip}{\extraspaces}
\addtolength{\abovedisplayshortskip}{\extraspace}
\addtolength{\belowdisplayshortskip}{\extraspace}}
\newcommand{\ee}{\end{equation}}
\newcommand{\ba}{\begin{eqnarray}
\addtolength{\abovedisplayskip}{\extraspaces}
\addtolength{\belowdisplayskip}{\extraspaces}
\addtolength{\abovedisplayshortskip}{\extraspace}
\addtolength{\belowdisplayshortskip}{\extraspace}}
\newcommand{\ea}{\end{eqnarray}}
\newcommand{\nonu}{\nonumber \\[.5mm]}
\newcommand{\A}{&\!\!\!}

\begin{document}
\thispagestyle{empty}
\setlength{\baselineskip}{6mm}
%
%\begin{flushright}
%{\tt hep-th/} \\
%March, 2022
%\end{flushright}
\vspace*{5mm}
\begin{center}
{\large\bf On commutator-based linearization of vector-spinor nonlinear \\[2mm]
supersymmetry and Rarita-Schwinger fields 
} \\[20mm]
{\sc Motomu Tsuda}
\footnote{
\tt e-mail: motomu.tsuda@gmail.com} 
\\[5mm]
{\it Aizu Hokurei High School \\
Aizuwakamatsu, Fukushima 965-0031, Japan} \\[20mm]
\begin{abstract}
We discuss the linearization of vector-spinor (spin-3/2) nonlinear supersymmetry (vsNLSUSY) transformations 
for both $N = 1$ and $N$-extended SUSY in flat spacetime 
based on the commutator algebra by using functionals (composites) of spin-3/2 Nambu-Goldstone (NG) fermions, 
which are expressed as simple products of powers of the spin-3/2 NG fermions 
and a fundamental determinant in the vsNLSUSY theory. 
We define basic component fields by means of those functionals in a linearized vsSUSY theory 
including spin-3/2 fields, general auxiliary ones and a $D$-term. 
The general forms of linear (rigid) vsSUSY transformations for the component fields 
are determined uniquely from the commutator-based linearization procedure. 
By considering appropriate recombinations of the functionals of the spin-3/2 NG fermions for $N = 1$ SUSY, 
we find that variations of the recombinations under the vsNLSUSY transformations 
include linear spin-1/2 SUSY ones of spin-(3/2, 1) fields with $U(1)$ gauge invariance. 
The spinorial gauge invariance of the Rarita-Schwinger action in the linearization process 
is also discussed together with the $U(1)$ gauge invariance. 
\\[5mm]
%
%\noindent
%PACS: 04.50.+h, 11.30.Pb, 12.60.Jv, 12.60.Rc, 12.10.-g \\[2mm]
%%
%\noindent
%Keywords: supersymmetry, superfield, composite unified theory 
%
\end{abstract}
\end{center}

\newpage

%\section{Introduction}

%\noindent
Rarita-Schwinger fields play important roles in the construction of the supergravity theory \cite{FNF,DZ} 
and massive spin-3/2 supermultiplets (for example, see \cite{Zi,BGLP}), etc. 
From the viewpoint of nonlinear realization of vector-spinor supersymmetry (vsNLSUSY) \cite{Ba}-\cite{PR2}, 
the non-manifest invariance of a (free) Rarita-Schwinger action in three dimensional spacetime 
under (rigid) vsSUSY transformations of the Rarita-Schwinger goldstinos 
was also shown by field redefinitions from the spin-3/2 goldstino model \cite{BS}. 
Moreover, NLSUSY general relativisitic theory \cite{Sh} 
expressed in terms of the vierbein field and spin-1/2 Nambu-Goldstone (NG) fermions, 
which was constructed as a spontaneously broken (rigid) SUSY model, 
is extended to vsNLSUSY one with spin-3/2 NG fermions \cite{ST1}. 

In order to investigate such theories including the spin-3/2 fermions in more detail, 
it is worth considering the relations between the vsNLSUSY theory and linear (L) vsSUSY ones 
as a wider framework with respect to spin structures for LSUSY 
from the viewpoint of the unitary representation of the vsSUSY algebra \cite{ST1a}. 
In addition, to study the vsSUSY may be useful for the considerations 
of higher-spin (consistent interaction) theories. 
In the case of spin-1/2 SUSY generated by the spinor parameters, 
the Volkov-Akulov NLSUSY theory \cite{VA} and LSUSY ones with $D$ terms 
indicating a spontaneously broken SUSY are related to each other (for example, see \cite{IK}-\cite{UZ}). 
The relation between NL and LSUSY (NL/LSUSY relation) gives deep insights into, 
e.g. the complicated structure of auxiliary fields, 
in particular in $N$-extended spin-1/2 LSUSY theories, 
\footnote{
The (most) general structure of the auxiliary fields in $N$ spin-1/2 LSUSY theories 
explicitly appears in a linearization based on the commutator algebra 
for spin-1/2 NLSUSY transformations \cite{MT1}. 
}
an origin of mass \cite{ST2} and a gauge coupling constant \cite{ST3}. 
A goldstino model for low-energy effective theories was also constructed 
by means of a nilpotent chiral superfield in Ref.\cite{KoS}. 

%%%%%%%%%%%%%%%%%%%%%%%%%%%%%%%%%%%%%%%%%%%%%%%%%%%%%%%%%%%%%%%%%%%%%%

The main building block of our constructions, the spin-3/2 NG fermions, 
does not carry any symmetries besides nonlinear realization of Baaklini symmetry \cite{Ba} 
and its ordinary SUSY subsymmetry and global $U(N)$ symmetry in the extended case. 
No any gauge $U(1)$ and/or spinorial Rarita-Schwinger symmetries are realized on this quantity, 
so any functional of these building blocks also cannot be a carrier of such gauge symmetries. 
The same concerns any invariant Lagrangian constructed out of such functionals. 
All terms "gauge transformations", "gauge parameters", "$U(1)$ gauge fields", etc. in this letter 
actually refer to some composite objects composed of the basic spin-3/2 and spin-1/2 NG fermions 
and constant parameters of rigid supersymmetries, like the scalar parameter $W_\zeta$ in Eq.(\ref{delta-v}) 
or the spinor parameter $X_\zeta$ in the discussion of the spinorial gauge invariance of the Rarita-Schwinger action 
in the last part of this letter. 
This is a crucial difference from the standard supersymmetric gauge theories 
where the genuine gauge invariances are present from the very beginning (alongside with rigid supersymmetry), 
and, after passing to the Wess-Zumino gauge \cite{WZ}, just compensating gauge transformations 
need to be added to the original supersymmetry variations in order to preserve the Wess-Zumino gauge. 

%%%%%%%%%%%%%%%%%%%%%%%%%%%%%%%%%%%%%%%%%%%%%%%%%%%%%%%%%%%%%%%%%%%%%%

While the vsSUSY is nonlinearly realized in Refs.\cite{Ba}-\cite{PR2}, 
the vsNL/vsLSUSY relation is not clear enough to our knowledge. 
In this letter, we focus on the vsNLSUSY theory constructed by Baaklini \cite{Ba} 
and discuss the linearization of the vsNLSUSY for $N = 1$ and $N$-extended SUSY 
by using a linearization procedure based on the commutator algebra \cite{MT1}. 
In the commutator-based linearization for the spin-1/2 NLSUSY theory, 
functional (composite) structures of spin-1/2 NG fermions in basic component fields for linearized spin-1/2 SUSY theories 
are manifest and LSUSY transformations with general auxiliary fields are determined uniquely and straightforwardly 
from the variations of the functionals under NLSUSY transformations. 
Hence, in this method of the linearization the complicated structure of auxiliary fields 
is explicitly shown also for vsLSUSY theories from the viewpoint of the compositeness of the NG fermions. 

Applying the commutator-based linearization procedure to the vsNLSUSY theory, 
we introduce in this letter bosonic and fermionic functionals of the spin-3/2 NG fermions 
as basic component fields in the linearized vsSUSY theory, 
which are represented as simple products of powers of the spin-3/2 NG fermions 
and a fundamental determinant in the vsNLSUSY theory. 
By using those functionals, we derive explicit and general forms of (rigid) vsLSUSY transformations 
of the component fields with spin-3/2 fields, general auxiliary ones and a $D$-term 
for spontaneously broken SUSY models. 

Furthermore, we consider appropriate recombinations of the functionals of the spin-3/2 NG fermions 
and study $U(1)$ gauge invariance in the linearized vsSUSY theory. 
From variations of the recombinations under the vsNLSUSY transformations 
we show at least that they include spin-1/2 LSUSY transformations among spin-(3/2, 1) fields with $U(1)$ gauge invariance, 
%%%%%%%%%%%%%%%%%%%%
which appear in the $N = 1$ gauge (3/2, 1) multiplet \cite{OS} 
%%%%%%%%%%%%%%%%%%%%
and in the massless limit of the massive spin-3/2 supermultiplets (e.g., in \cite{Zi,BGLP}). 
%%%%%%%%%%%%%%%%%%%%
The spinorial gauge invariance of the Rarita-Schwinger action in the linearization process 
is also discussed together with the $U(1)$ gauge invariance. 
%%%%%%%%%%%%%%%%%%%%

%%%%%%%%%%%%%%%%%%%%%%%%%%%%%%%%%%%%%%%%%%%%%%%%%%%%%%%%%%%%
Let us start with the vsSUSY algebra in the Baaklini's vsNLSUSY model \cite{Ba} for $N$-extended SUSY, 
\ba
\A \A 
\{ Q^{(i)a}_\alpha, Q^{(j)b}_\beta \} = i \delta^{(i)(j)} \epsilon^{abcd} (\gamma_5 \gamma_c C)_{\alpha \beta} P_d, 
\nonu
\A \A 
[Q^{(i)a}_\alpha, P^b] = 0, 
\nonu
\A \A 
[Q^{(i)a}_\alpha, J^{bc}] = {1 \over 2} (\sigma^{bc})_\alpha{}^\beta Q^{(i)a}_\beta 
+ i (\eta^{ab} Q^{(i)c}_\alpha - \eta^{ac} Q^{(i)b}_\alpha), 
\label{vsNLSUSYalg}
\ea
where $Q^{(i)a}_\alpha$ are (Majorana) vector-spinor generators, 
while $P^a$ and $J^{ab}$ are translational and Lorentz generators of the Poincar\`{e} group. 
\footnote{
Minkowski spacetime indices are denoted by $a, b, \cdots = 0, 1, 2, 3$ 
and the internal indices are $(i),(j), \cdots = 1, 2, \cdots, N$. 
The gamma matrices satisfy $\{ \gamma^a, \gamma^b \} = 2 \eta^{ab}$ 
with the Minkowski spacetime metric $\eta^{ab} = {\rm diag}(+,-,-,-)$ 
and $\displaystyle{\sigma^{ab} = {i \over 2}[\gamma^a, \gamma^b]}$ is defined. 
The charge conjugation matrix is $C = -i \gamma_0 \gamma_2$. 
}
The nonlinear realization of the vsSUSY algebra (\ref{vsNLSUSYalg}) is given 
by defining the vsNLSUSY transformations in terms of spin-3/2 NG (Majorana) fermions $\psi^{(i)a}$ \cite{Ba} as 
%%%%%%%%%%%%%%%%%%%%%%%%%%%%%%%%%%%%%%%%%%%%%%%%%%
%
\be
\delta_\zeta \psi^{(i)a} 
%\A = \A {1 \over \kappa} \zeta^a + \kappa \epsilon^{bcde} \bar\zeta_b \gamma_5 \gamma_c \psi_d \partial_e \psi^a, 
%\nonu
= {1 \over \kappa} \zeta^{(i)a} + \xi^b \partial_b \psi^{(i)a}, 
\label{vsNLSUSY}
\ee
where $\kappa$ means a dimensional constant whose dimension is (mass)$^{-2}$, 
$\zeta^{(i)a}$ are (constant) vsSUSY transformation parameters 
and $\xi^a %= \kappa \epsilon^{cdeb} \bar\zeta_c \gamma_5 \gamma_d \psi_e 
= \kappa \epsilon^{abcd} \bar\psi^{(i)}_b \gamma_5 \gamma_c \zeta^{(i)}_d$. 
The commutator algebra for the NLSUSY transformations (\ref{vsNLSUSY}) is closed off-shell as 
\be
[\delta_{\zeta_1}, \delta_{\zeta_2}] = \delta_P(\Xi^a), 
\label{vsNLSUSYcomm}
\ee
where $\delta_P(\Xi^a)$ means a translation with parameters 
$\Xi^a = 2 \epsilon^{abcd} \bar\zeta^{(i)}_{1b} \gamma_5 \gamma_c \zeta^{(i)}_{2d}$. 
By introducing a fundamental determinant, 
\be
\vert w_{3/2} \vert = \det \{ (w_{3/2})^a{}_b \} = \det \{ \delta^a_b + (t_{3/2})^a{}_b \} 
= \det(\delta^a_b + \kappa^2 \epsilon^{acde} \bar\psi^{(i)}_c \gamma_5 \gamma_d \partial_b \psi^{(i)}_e), 
%= 1 + t^a{}_a + {1 \over 2!}(t^a{}_a t^b{}_b - t^a{}_b t^b{}_a) 
\label{determinant}
\ee
a vsNLSUSY action is defined as 
\be
S_{\rm vsNLSUSY} = -{1 \over {2 \kappa^2}} \int d^4 x \vert w_{3/2} \vert, 
%\nonu
%\A = \A - {1 \over {2 \kappa^2}} 
%\left\{ 1 + t^a{}_a + {1 \over 2!} (t^a{}_a t^b{}_b - t^a{}_b t^b{}_a)
%\right. 
%\nonu
%\A \A 
%\left. 
%- {1 \over 3!} \epsilon_{abcd} \epsilon^{efgd} t^a{}_e t^b{}_f t^c{}_g 
%- {1 \over 4!} \epsilon_{abcd} \epsilon^{efgd} t^a{}_e t^b{}_f t^c{}_g t^d{}_h 
%\right\}, 
\label{vsNLSUSYaction}
\ee
due to $\delta_\zeta \vert w_{3/2} \vert = \partial_a (\xi^a \vert w_{3/2} \vert)$. 

Here let us consider bosonic and fermionic functionals which are composed of products of 
$(\psi^{(i)a})^{2(n-1)}$-terms ($n = 1, 2, \cdots$) and $\vert w_{3/2} \vert$, 
or those of $(\psi^{(i)a})^{2n-1}$-ones and $\vert w_{3/2} \vert$ as follows; 
namely, the bosonic functionals are 
\ba
\A \A 
b^{(i)(j)(k)(l) \cdots (m)(n)a}{}_A{}^{bc}{}_B{}^{d \cdots e}{}_C{}^f \left( (\psi^{(i)a})^{2(n-1)} \vert w_{3/2} \vert \right) 
\nonu
\A \A 
\hspace{1cm} = \kappa^{2n-3} \bar\psi^{(i)a} \gamma_A \psi^{(j)b} \bar\psi^{(k)c} \gamma_B \psi^{(l)d} 
\cdots \bar\psi^{(m)e} \gamma_C \psi^{(n)f} \vert w_{3/2} \vert 
\label{bosonic}
\ea
which mean 
\ba
\A \A 
b = \kappa^{-1} \vert w_{3/2} \vert, \ \ b^{(i)(j)a}{}_A{}^b = \kappa \bar\psi^{(i)a} \gamma_A \psi^{(j)b} \vert w_{3/2} \vert, 
\nonu
\A \A 
b^{(i)(j)(k)(l)a}{}_A{}^{bc}{}_B{}^d = \kappa^3 \bar\psi^{(i)a} \gamma_A \psi^{(j)b} \bar\psi^{(k)c} \gamma_B \psi^{(l)d} \vert w_{3/2} \vert, 
\ \cdots, 
\ea
etc., while the fermionic ones are 
\ba
\A \A 
f^{(i)(j)(k)(l)(m) \cdots (n)(p)ab}{}_A{}^{cd}{}_B{}^{e \cdots f}{}_C{}^g \left( (\psi^{(i)a})^{2n-1} \vert w_{3/2} \vert \right) 
\nonu
\A \A 
\hspace{1cm} = \kappa^{2(n-1)} \psi^{(i)a} \bar\psi^{(j)b} \gamma_A \psi^{(k)c} \bar\psi^{(l)d} \gamma_B \psi^{(m)e} 
\cdots \bar\psi^{(n)f} \gamma_C \psi^{(p)g} \vert w_{3/2} \vert 
\label{fermionic}
\ea
which express 
\be
f^{(i)a} = \psi^{(i)a} \vert w_{3/2} \vert, 
\ \ f^{(i)(j)(k)ab}{}_A{}^c = \kappa^2 \psi^{(i)a} \bar\psi^{(j)b} \gamma_A \psi^{(k)c} \vert w_{3/2} \vert, 
%\ \ f^{ab}{}_A{}^{cd}{}_B{}^e = \kappa^4 \psi^a \bar\psi^b \gamma_A \psi^c \bar\psi^d \gamma_B \psi^e \vert w_{3/2} \vert, 
\ \cdots, 
\ee
etc. 
In these functionals, (Lorentz) indices $A, B, \cdots$ are used as ones for a basis of $\gamma$ matrices, 
i.e. $\gamma_A = {\bf 1}, i \gamma_5, i \gamma_a, \gamma_5 \gamma_a$ or $\sqrt{2} i \sigma_{ab}$. 
%
%$f^a$ give the leading order of superchages $Q^a$ and 
%
The definition of the functionals (\ref{bosonic}) and (\ref{fermionic}) terminate 
with $n = 8N + 1$ and $n = 8N$, respectively, because $(\psi^{(i)a})^{16N+1} = 0$. 
%
%with $n = 9$ and $n = 8$, respectively, because $(\psi^a)^{17} = 0$. 
We also note that the functionals (\ref{bosonic}) and (\ref{fermionic}) satisfy the commutator algebra (\ref{vsNLSUSYcomm}) 
under the vsNLSUSY transformations (\ref{vsNLSUSY}). 
\footnote{
In the same way as in the spin-1/2 NLSUSY theory, 
it can be proved that every Lorentz-tensor (or scalar) functionals composed of $\psi^{(i)a}$ and their derivative terms 
satisfies the same commutator algebra (\ref{vsNLSUSYcomm}) (for example, see \cite{ST4}). 
}

Then, the variations of the functionals (\ref{bosonic}) and (\ref{fermionic}) 
under the vsNLSUSY transformations (\ref{vsNLSUSY}) become 
\ba
\A \A 
\delta_\zeta b^{(i)(j)(k)(l) \cdots (m)(n)a}{}_A{}^{bc}{}_B{}^{d \cdots e}{}_C{}^f 
\nonu
\A \A 
\hspace{8mm} = \kappa^{2(n-2)} \left[ \left\{ \left( 
\bar\zeta^{(i)a} \gamma_A \psi^{(j)b} + \bar\psi^{(i)a} \gamma_A \zeta^{(j)b} \right) 
\bar\psi^{(k)c} \gamma_B \psi^{(l)d} \cdots \bar\psi^{(m)e} \gamma_C \psi^{(n)f} 
+ \cdots \right\} \vert w_{3/2} \vert \right. 
\nonu
\A \A 
\hspace{1.1cm} \left. + \kappa \partial_g \left( \xi^g \bar\psi^{(i)a} \gamma_A \psi^{(j)b} \bar\psi^{(k)c} \gamma_B \psi^{(l)d} 
\cdots \bar\psi^{(m)e} \gamma_C \psi^{(n)f} \vert w_{3/2} \vert \right) \right], 
\label{variation1}
\\[1mm]
\A \A 
\delta_\zeta f^{(i)(j)(k)(l)(m) \cdots (n)(p)ab}{}_A{}^{cd}{}_B{}^{e \cdots f}{}_C{}^g
\nonu
\A \A 
\hspace{8mm} = \kappa^{2n-3} \left[ \left\{ \zeta^{(i)a} \bar\psi^{(j)b} \gamma_A \psi^{(k)c} 
\bar\psi^{(l)d} \gamma_B \psi^{(m)e} \cdots \bar\psi^{(n)f} \gamma_C \psi^{(p)g} 
\right. \right. 
\nonu
\A \A 
\hspace{1.1cm} \left. + \psi^{(i)a} \left( \bar\zeta^{(j)b} \gamma_A \psi^{(k)c} + \bar\psi^{(j)b} \gamma_A \zeta^{(k)c} \right) 
\bar\psi^{(l)d} \gamma_B \psi^{(m)e} \cdots \bar\psi^{(n)f} \gamma_C \psi^{(p)g} + \cdots \right\} \vert w_{3/2} \vert 
\nonu
\A \A 
\hspace{1.1cm} \left. + \kappa \partial_h \left( \xi^h \psi^{(i)a} \bar\psi^{(j)b} \gamma_A \psi^{(k)c} \bar\psi^{(l)d} \gamma_B \psi^{(m)e} 
\cdots \bar\psi^{(n)f} \gamma_C \psi^{(p)g} \vert w_{3/2} \vert \right) \right]. 
\label{variation2}
\ea
%
%by using the variations of $\vert w \vert$, i.e. $\delta_\zeta \vert w \vert = \partial_a (\xi^a \vert w \vert)$. 
%
These variations (\ref{variation1}) and (\ref{variation2}) indicate that the bosonic and fermionic functionals 
in Eqs.(\ref{bosonic}) and (\ref{fermionic}) are linearly exchanged with each other. 
%Indeed, the functionals (\ref{bosonic}) and (\ref{fermionic}) lead to LSUSY transformations of component fields 
%in (massless) vector linear supermultiplets with general auxiliary fields, 
%%prior to transforming to gauge supermultiplets, 
%e.g. in the case for $N = 2$ SUSY in two-dimensional spacetime \cite{ST3,ST4}. 
%They are also proper functionals in order to study the NL/LSUSY relations for vector supermultiplets 
%in extended SUSY because of the vector components in the functionals $b^i{}_A{}^j$ for $N \ge 2$ SUSY. 

%\noindent
%(Note) \\
%$\gamma^A = {\bf 1}, -i \gamma_5, -i \gamma^a, -\gamma_5 \gamma^a \ {\rm or} \ -\sqrt{2} i \sigma^{ab}$ \\
%The sign factor $\varepsilon$ is $\varepsilon = +1$ for $\gamma_A = {\bf 1}, i \gamma_5, \gamma_5 \gamma_a$ 
%and $\varepsilon = -1$ for $\gamma_A = i \gamma_a, \sqrt{2} i \sigma_{ab}$. 

Next we explain the commutator-based linearization of the vsNLSUSY 
by defining basic component fields in vsLSUSY multiplets as 
\ba
\A \A 
D = b = \kappa^{-1} \vert w_{3/2} \vert, 
\nonu
\A \A 
\lambda^{(i)a} = f^{(i)a}(\psi) = \psi^{(i)a} \vert w_{3/2} \vert, 
\nonu
\A \A 
M^{(i)(j)a}{}_A{}^b = \alpha_{1A} b^{(i)(j)a}{}_A{}^b(\psi) = \alpha_{1A} \kappa \bar\psi^{(i)a} \gamma_A \psi^{(j)b} \vert w_{3/2} \vert, 
\nonu
\A \A 
\Lambda^{(i)(j)(k)ab}{}_A{}^c = \alpha_{2A} f^{(i)(j)(k)ab}{}_A{}^c(\psi) 
= \alpha_{2A} \kappa^2 \psi^{(i)a} \bar\psi^{(j)b} \gamma_A \psi^{(k)c} \vert w_{3/2} \vert
\nonu
\A \A 
C^{(i)(j)(k)(l)a}{}_A{}^{bc}{}_B{}^d = \alpha_{3AB} b^{(i)(j)(k)(l)a}{}_A{}^{bc}{}_B{}^d(\psi) 
\nonu
\A \A 
\hspace{3cm} = \alpha_{3AB} \kappa^3 \bar\psi^{(i)a} \gamma_A \psi^{(j)b} \bar\psi^{(k)c} \gamma_B \psi^{(l)d} \vert w_{3/2} \vert, 
\label{comp}
\ea
%
%
%\ba
%\A \A 
%D = b \left( \vert w \vert \right), 
%\\
%\A \A 
%\lambda^{(i)a} = f^{(i)a} \left( \psi^{(i)a} \vert w \vert \right), 
%\\
%\A \A 
%M^{(ij)a}{}_A{}^b = \alpha_{1A} b^{(ij)a}{}_A{}^b \left( (\psi^{(i)a})^2 \vert w \vert \right), 
%\\
%\A \A 
%\Lambda^{(ijk)ab}{}_A{}^c = \alpha_{2A} f^{(ijk)ab}{}_A{}^c \left( (\psi^{(i)a})^3 \vert w \vert \right)
%\\
%\A \A 
%C^{(ijkl)a}{}_A{}^{bc}{}_B{}^d 
%= \alpha_{3AB} b^{(ijkl)a}{}_A{}^{bc}{}_B{}^d \left( (\psi^{(i)a})^4 \vert w \vert \right)
%\label{comp}
%\ea
%
$\cdots$ etc., 
%%%%%%%%%%%%%%%%%%%%%%%%%%%%%%%%%%%%%%%%%%%%%%%%%%%%%%%%%%%%%%%%%%%%%%%%%%%%%%%%
%
where ($\alpha_{1A}, \alpha_{2A}, \alpha_{3AB}$) are arbitrary constants. 
%
%where values of the constants ($\alpha_{1A}, \alpha_{2A}, \alpha_{3AB}$) should be determined 
%from the definition of a fundamental action in a vsLSUSY theory 
%and the invariance of the action under vsLSUSY transformations of the component fields. 
%
%%%%%%%%%%%%%%%%%%%%%%%%%%%%%%%%%%%%%%%%%%%%%%%%%%%%%%%%%%%%%%%%%%%%%%%%%%%%%%%%
%
We notice that the definition (\ref{comp}) (and also the forms of the functionals in Eqs.(\ref{bosonic}) to (\ref{variation2})) 
are just (the simplest) ansatz, which can include the derivatives of the spin-3/2 NG fermions 
besides the hidden ones in $\vert w_{3/2} \vert$. 
In the linearization of the VA NLSUSY theory, the derivatives of the NG fermion are considered 
in functionals to determine basic fields in LSUSY theories according to the general prescription in Ref.\cite{IK}. 
In fact, we reconsidered in Ref.\cite{ST-hd} that basic fields in the $N = 1$ scalar supermultiplet 
are expanded in terms of the NG fermion together with those kind of the (higher) derivatives in the SUSY invariant way 
and then the LSUSY action is related to the VA action with (apparently pathological) higher derivative terms, 
which reduces to the (standard) VA one without their derivative terms by a field redefinition. 
Therefore, we expect that such a universal structure with respect to the derivatives of the NG fermion 
also exists in the case of the linearization of vsNLSUSY. 

%%%%%%%%%%%%%%%%%%%%%%%%%%%%%%%%%%%%%%%%%%%%%%%%%%%%%%%%%%%%%%%%%%%%%%%%%%%%%%%%

According to the variations (\ref{variation1}) and (\ref{variation2}), 
vsLSUSY transformations for the components $D$ and $\lambda^{(i)a}$ in Eq.(\ref{comp}) 
under the vsNLSUSY transformations (\ref{vsNLSUSY}) are unambigiously determined as follows: 
\ba
\A \A
\delta_\zeta D = - \epsilon^{abcd} \bar\zeta^{(i)}_a \gamma_5 \gamma_b \partial_c \lambda^{(i)}_d, 
\label{variationD}
\\
\A \A
\delta_\zeta \lambda^{(i)a} = D \zeta^{(i)a} 
- {1 \over {4 \alpha_{1A}}} \varepsilon_A \epsilon^{bcde} \partial_b M^{(i)(j)a}{}_{Ac} \gamma^A \gamma_5 \gamma_d \zeta^{(j)}_e. 
\label{variation-llambda}
\ea
In Eq.(\ref{variation-llambda}), $\varepsilon$ ($\varepsilon_A$ etc.) 
means a sign factor which appear from the relation 
$\bar\psi^{(i)a} \gamma_A \psi^{(j)b} = \varepsilon_A \bar\psi^{(j)b} \gamma_A \psi^{(i)a}$ 
\footnote
{The sign factor $\varepsilon$ is $\varepsilon = +1$ for $\gamma_A = {\bf 1}, i \gamma_5, \gamma_5 \gamma_a$ 
and $\varepsilon = -1$ for $\gamma_A = i \gamma_a, \sqrt{2} i \sigma_{ab}$. 
}
and it is also used in the same meaning below. 

As for the other components in Eq.(\ref{comp}), 
arguments based on the commutator algebra (\ref{vsNLSUSYcomm}) are necessary 
in order to determine the explicit forms of vsLSUSY transformations. 
For instance, let us consider variations of $M^{(i)(j)a}{}_A{}^b$ under the vsNLSUSY transformations (\ref{vsNLSUSY}), i.e. 
\be
{1 \over \alpha_{1A}} \delta_\zeta M^{(i)(j)a}{}_A{}^b 
= \bar\zeta^{(i)a} \gamma_A \lambda^{(j)b} + \bar\lambda^{(i)a} \gamma_A \zeta^{(j)b} 
- \kappa^2 \partial_c \left( \epsilon^{cdef} \bar\zeta^{(k)}_d \gamma_5 \gamma_e \psi^{(k)}_f 
\bar\psi^{(i)a} \gamma_A \psi^{(j)b} \vert w_{3/2} \vert \right). 
%
%+ {1 \over {4 \alpha_{2B}}} \varepsilon_A \epsilon^{cdef} 
%\bar\zeta^{(k)}_c \gamma_5 \gamma_d \gamma^B \gamma_A \partial_e \Lambda^{(i)(j)(k)ab}{}_{Bf}. 
\label{variation-M0}
\ee
Since the variations (\ref{variation-M0}) include a problem for deformations of the $(\psi^{(i)a})^3$-functionals in the last terms, 
we examine two supertransfomations of $\lambda^{(i)a}$, 
\ba
\delta_{\zeta_1} \delta_{\zeta_2} \lambda^{(i)a} 
\A = \A \left( \delta_{\zeta_1} D \right) \zeta_2^{(i)a} 
- {1 \over {4 \alpha_{1A}}} \varepsilon_A \epsilon^{bcde} \partial_b 
\left( \delta_{\zeta_1} M^{(i)(j)a}{}_{Ac} \right) \gamma^A \gamma_5 \gamma_d \zeta^{(j)}_{2e} 
\nonu
\A = \A \epsilon^{bcde} \bar\zeta^{(j)}_{1c} \gamma_5 \gamma_d \zeta^{(j)}_{2e} \partial_b \lambda^{(i)a}
+ \left[ (1 \leftrightarrow 2)\ {\rm symmetric\ terms\ of}\ \partial_a \lambda^{(i)}_b \right] 
\nonu
\A \A 
\!\!\!\!\! - {1 \over 16} \kappa^2 \varepsilon_A \epsilon^{bcde} \epsilon^{fghi} 
\partial_b \partial_f \left( \bar\zeta^{(k)}_{1g} \gamma_B \zeta^{(j)}_{2e} 
\gamma_A \gamma_5 \gamma_d \gamma^B \gamma_5 \gamma_h \psi^{(k)}_i 
\bar\psi^{(i)a} \gamma_A \psi^{(j)}_c \vert w_{3/2} \vert \right). 
\label{twosuper-lambda}
\ea
Eq.(\ref{twosuper-lambda}) satisfies the commutator algebra (\ref{vsNLSUSYcomm}) on the vsNLSUSY phase 
and the last terms of Eq.(\ref{twosuper-lambda}) also vanish in the commutator algebra. 
Therefore, those terms are symmetric under exchanging the indices $1$ and $2$ 
in the vsSUSY transformation parameters ($\zeta^{(k)}_{1g}$, $\zeta^{(j)}_{2e}$). 
This fact means that if the $\psi^{(k)}_f$ and $\psi^{(j)b}$ in the last terms of Eq.(\ref{variation-M0}) 
go into bilinear forms $\bar\psi^{(j)b} \gamma^B \psi^{(k)}_f$ with respect to the internal indices $(j)$ and $(k)$ 
by using a Fierz transformation and if those terms are expressed 
by means of the fermionic components $\Lambda^{(i)(j)(k)ab}{}_A{}^c$ in Eq.(\ref{comp}) as 
\ba
\A \A - \kappa^2 \partial_c \left( \epsilon^{cdef} \bar\zeta^{(k)}_d \gamma_5 \gamma_e \psi^{(k)}_f 
\bar\psi^{(i)a} \gamma_A \psi^{(j)b} \vert w_{3/2} \vert \right) 
\nonu
\A \A \hspace{1cm} = {1 \over 4} \kappa^2 \varepsilon_A \partial_c \left( \epsilon^{cdef} 
\bar\zeta^{(k)}_d \gamma_5 \gamma_e \gamma_B \gamma_A \psi^{(i)a} \bar\psi^{(j)b} \gamma^B \psi^{(k)}_f \vert w_{3/2} \vert \right) 
\nonu
\A \A \hspace{1cm} = {1 \over {4 \alpha_{2B}}} \varepsilon_A \epsilon^{cdef} 
\bar\zeta^{(k)}_c \gamma_5 \gamma_d \gamma^B \gamma_A \partial_e \Lambda^{(i)(j)(k)ab}{}_{Bf}, 
\ea
prior to the two supertransfomations (\ref{twosuper-lambda}), 
then we can straightforwardly confirm the vanishment of the last terms of Eq.(\ref{twosuper-lambda}) 
in the commutator algebra on the vsLSUSY phase. 
%
%This means that the vanishments of the last terms of Eq.(\ref{twosuper-lambda}) in Eq.(\ref{NLSUSYcomm2}) 
%can be confirmed straightforwardly when the $\psi^k$ and $\psi^j$ go into bilinear forms 
%$\bar\psi^j \gamma_A \psi^k$ in the last terms of Eqs.(\ref{v-M}) and (\ref{twosuper-lambda}), 
%which have the same indices as the spinor transformation parameters 
%and reflect the symmetries of the $\bar\zeta_1^k \gamma_B \zeta_2^j$. 
%
%Since these two supertransformations 
%satisfy Eq.(\ref{NLSUSYcomm2}), 
%the last terms of Eq.(\ref{twosuper-lambda}) which vanish in the commutation relations 
%%(obtained through $\delta_\zeta M^i{}_A{}^j$) 
%have to be symmetric under exchanging the indices $1$ and $2$ 
%of the spinor transformation parameters ($\zeta_1^k$, $\zeta_2^j$). 
%This means that the vanishments of the last terms of Eq.(\ref{twosuper-lambda}) in Eq.(\ref{NLSUSYcomm2}) 
%can be confirmed straightforwardly when the $\psi^k$ and $\psi^j$ go into bilinear forms 
%$\bar\psi^j \gamma_A \psi^k$ in the last terms of Eqs.(\ref{v-M}) and (\ref{twosuper-lambda}), 
%which have the same indices as the spinor transformation parameters 
%and reflect the symmetries of the $\bar\zeta_1^k \gamma_B \zeta_2^j$. 
%
Thus the vsLSUSY transformations of $M^{(i)(j)a}{}_A{}^b$ are uniquely determined as 
\be
{1 \over \alpha_{1A}} \delta_\zeta M^{(i)(j)a}{}_A{}^b 
= \bar\zeta^{(i)a} \gamma_A \lambda^{(j)b} + \bar\lambda^{(i)a} \gamma_A \zeta^{(j)b} 
+ {1 \over {4 \alpha_{2B}}} \varepsilon_A \epsilon^{cdef} 
\bar\zeta^{(k)}_c \gamma_5 \gamma_d \gamma^B \gamma_A \partial_e \Lambda^{(i)(j)(k)ab}{}_{Bf}. 
\label{variationM}
\ee

In the same way, we can also determine the form of vsLSUSY transformations for $\Lambda^{(i)(j)(k)ab}{}_A{}^c$ in Eq.(\ref{comp}) 
by examining variations of $\Lambda^{(i)(j)(k)ab}{}_A{}^c$ under the vsNLSUSY transformations (\ref{vsNLSUSY}), 
\ba
\A \A 
{1 \over \alpha_{2A}} \delta_\zeta \Lambda^{(i)(j)(k)ab}{}_A{}^c 
= {1 \over \alpha_{1A}} M^{(j)(k)b}{}_A{}^c \zeta^{(i)a} 
- {1 \over {4 \alpha_{1B}}} (\varepsilon_A M^{(k)(i)c}{}_B{}^a \gamma^B \gamma_A \zeta^{(j)b} 
+ M^{(j)(i)b}{}_B{}^a \gamma^B \gamma_A \zeta^{(k)c}) 
\nonu
\A \A 
\hspace{2.5cm} 
- {1 \over 4} \kappa^3 \partial_d \left( \epsilon^{defg} \gamma^B \gamma_5 \gamma_f \zeta^{(l)}_g 
\bar\psi^{(l)}_e \gamma_B \psi^{(i)a} \bar\psi^{(j)b} \gamma_A \psi^{(k)c} \vert w_{3/2} \vert \right), 
\label{variation-lambda0}
\ea
and two supertransformations of $M^{(i)(j)a}{}_A{}^b$, 
\ba
\A \A 
{1 \over \alpha_{1A}} \delta_{\zeta_1} \delta_{\zeta_2} M^{(i)(j)a}{}_A{}^b 
= \bar\zeta_2^{(i)a} \gamma_A \left( \delta_{\zeta_1} \lambda^{(j)b} \right) 
+ \left( \delta_{\zeta_1} \bar\lambda^{(i)a} \right) \gamma_A \zeta_2^{(j)b} 
\nonu
\A \A 
\hspace{3.7cm} + {1 \over {4 \alpha_{2B}}} \varepsilon_A \epsilon^{cdef} 
\bar\zeta^{(k)}_{2c} \gamma_5 \gamma_d \gamma^B \gamma_A \partial_e 
\left( \delta_{\zeta_1} \Lambda^{(i)(j)(k)ab}{}_{Bf} \right) 
\nonu
\A \A 
= \epsilon^{cdef} \bar\zeta^{(k)}_{1d} \gamma_5 \gamma_e \zeta^{(k)}_{2f} \partial_c M^{(i)(j)a}{}_A{}^b 
+ \left[ (1 \leftrightarrow 2)\ {\rm symmetric\ terms\ of}\ \partial_a M^{(i)(j)b}{}_A{}^c \right] 
\nonu
\A \A 
\hspace{5mm} + {1 \over 16} \kappa^3 \partial_c \partial_g \left( \varepsilon_A \epsilon^{cdef} \epsilon^{ghij} 
\bar\zeta^{(k)}_{2d} \gamma_5 \gamma_e \gamma^B \gamma_A \gamma^C \gamma_5 \gamma_i \zeta^{(l)}_{1j} 
\bar\psi^{(l)}_h \gamma_C \psi^{(i)a} 
\bar\psi^{(j)b} \gamma_B \psi^{(k)}_f \vert w_{3/2} \vert \right). 
\label{twosuper-M}
\ea
We refer to the internal indices $(k)$ and $(l)$ for the vsSUSY transformation parameters ($\zeta^{(k)}_{2d}$, $\zeta^{(l)}_{1j}$) 
in the last terms of Eq.(\ref{twosuper-M}) which satisfies the commutator algebra (\ref{vsNLSUSYcomm}) on the vsNLSUSY phase. 
Then, the vsLSUSY transformations for $\Lambda^{(i)(j)(k)ab}{}_A{}^c$ are uniquely determined 
by transforming the $\psi^{(l)}_e$ and $\psi^{(k)c}$ in the last terms of Eq.(\ref{variation-lambda0}) 
into bilinear forms by using a Fierz transformation prior to Eq.(\ref{twosuper-M}) 
and by expressing Eq.(\ref{variation-lambda0}) in terms of the bosonic components 
$C^{(i)(j)(k)(l)a}{}_A{}^{bc}{}_B{}^d$ in Eq.(\ref{comp}) as 
\ba
%\A \A 
%\delta_\zeta D = - \epsilon^{abcd} \bar\zeta^{(i)}_a \gamma_5 \gamma_b \partial_c \lambda^{(i)}_d, 
%\\
%\A \A 
%\delta_\zeta \lambda^{(i)a} = D \zeta^{(i)a} 
%- {1 \over {4 \alpha_{1A}}} \varepsilon_A \epsilon^{bcde} \partial_b M^{(i)(j)a}{}_{Ac} \gamma^A \gamma_5 \gamma_d \zeta^{(j)}_e, 
%\\
%\A \A 
%{1 \over \alpha_{1A}} \delta_\zeta M^{(i)(j)a}{}_A{}^b 
%= \bar\zeta^{(i)a} \gamma_A \lambda^{(j)b} + \bar\lambda^{(i)a} \gamma_A \zeta^{(j)b} 
%+ {1 \over {4 \alpha_{2B}}} \varepsilon_A \epsilon^{cdef} 
%\bar\zeta^{(k)}_c \gamma_5 \gamma_d \gamma^B \gamma_A \partial_e \Lambda^{(i)(j)(k)ab}{}_{Bf},  
%\\
{1 \over \alpha_{2A}} \delta_\zeta \Lambda^{(i)(j)(k)ab}{}_A{}^c 
\A = \A {1 \over \alpha_{1A}} M^{(j)(k)b}{}_A{}^c \zeta^{(i)a} 
- {1 \over {4 \alpha_{1B}}} (\varepsilon_A M^{(k)(i)c}{}_B{}^a \gamma^B \gamma_A \zeta^{(j)b} 
+ M^{(j)(i)b}{}_B{}^a \gamma^B \gamma_A \zeta^{(k)c}) 
\nonu
\A \A 
%\hspace{2.5cm} 
+ {1 \over {16 \alpha_{3ACB}}} \varepsilon_C \varepsilon'_{ACB} \epsilon^{defg} 
\partial_d C^{(i)(j)(k)(l)aCbc}{}_{ACBe} \gamma^B \gamma_5 \gamma_f \zeta^{(l)}_g. 
\label{variation-lambda}
\ea
%
%$\cdots$ etc. 
The vsLSUSY transformations (\ref{variation-lambda}) also satisfy the commutator algebra (\ref{vsNLSUSYcomm}) 
on the vsLSUSY phase. 

When we further examine variations of the component fields for higher order functionals of $\psi^{(i)a}$, 
e.g. those of $C^{(i)(j)(k)(l)a}{}_A{}^{bc}{}_B{}^d$ under the vsNLSUSY transformations, 
additional problems for deformations of the higher order functionals of $\psi^{(i)a}$ 
appear in order to define vsLSUSY transformations for these components. 
However, those problems are also solved by examining two supertransformations 
of the components under vsNLSUSY phase (see \cite{MT1} for details). 
As for the degrees of freedom of the general bosonic and fermionic components, 
they should be balanced for each $N$ vsLSUSY theory from the same arguments 
as in the linearization of the spin-1/2 NLSUSY theory \cite{MT2} (in two dimensional spacetime), 
in which their degrees of freedom are determined by means of the functional structure 
of spin-1/2 NG fermions for the components. 
%%%%%%%%%%%%%%%%%%%%%%%%%%%%%%%%%%%%%%%%
%
In particular, the relations (identities) through Fierz transformations of $\psi^a$ 
give many constraints for the general auxiliary fields 
($\Lambda^{(i)(j)(k)ab}{}_A{}^c$, $C^{(i)(j)(k)(l)a}{}_A{}^{bc}{}_B{}^d$, $\cdots$) in the vsNLSUSY phase 
and the problem for counting the total degrees of freedom for the whole component fields 
for the vsLSUSY multiplet remains to be solved. 

%%%%%%%%%%%%%%%%%%%%%%%%%%%%%%%%%%%%%%%%

From now on let us consider for $N = 1$ SUSY how the above vsLSUSY transformations in the linearized vsSUSY theory 
include {\it spin-1/2 LSUSY} transformations among spin-(3/2, 1) fields with $U(1)$ gauge invariance, 
which appear in the massless limit of the massive spin-3/2 supermultiplets (e.g., in \cite{Zi,BGLP}). 
%%%%%%%%%%%%%%%%%%%%%%%%%%%%%%%%%%%%%%%%%%%%%%%%%%%%%%%%%%%%%%%%%%%%%%%%%%%%%%%%
%
%Concerning an action in the vsLSUSY theory, here we consider the following general basic fields 
%for $N = 1$ SUSY in the linear multiplet (\ref{comp}) for simplicity, 
%
The general basic fields for $N = 1$ SUSY in the linear multiplet (\ref{comp}) are 
\ba
\A \A 
D = b = \kappa^{-1} \vert w_{3/2} \vert, 
\nonu
\A \A 
\lambda^a = f^a(\psi) = \psi^a \vert w_{3/2} \vert, 
\nonu
\A \A 
M^{ab} = M^{(ab)} = \alpha_{11} b^{(ab)}(\psi) = \alpha_{11} \kappa \bar\psi^a \psi^b \vert w_{3/2} \vert, 
\nonu
\A \A 
M_5{}^{ab} = M_5{}^{(ab)} = \alpha_{12} b_5{}^{(ab)}(\psi) = \alpha_{12} i \kappa \bar\psi^a \gamma_5 \psi^b \vert w_{3/2} \vert, 
\nonu
\A \A 
M^a{}_c{}^b = M^{[a}{}_c{}^{b]} 
= \alpha_{13} b^{[a}{}_c{}^{b]}(\psi) = \alpha_{13} i \kappa \bar\psi^a \gamma_c \psi^b \vert w_{3/2} \vert, 
\nonu
\A \A 
M_5{}^a{}_c{}^b = M_5{}^{(a}{}_c{}^{b)} = \alpha_{14} b_5{}^{(a}{}_c{}^{b)}(\psi) = \alpha_{14} \kappa \bar\psi^a \gamma_5 \gamma_c \psi^b \vert w_{3/2} \vert, 
\nonu
\A \A 
M^a{}_{cd}{}^b = M^{[a}{}_{cd}{}^{b]} = \alpha_{15} b^{[a}{}_{cd}{}^{b]}(\psi) = \sqrt{2} \alpha_{15} i \kappa \bar\psi^a \sigma_{cd} \psi^b \vert w_{3/2} \vert, 
\nonu
\A \A 
\Lambda^{abc} = \Lambda^{a(bc)} = \alpha_{21} f^{a(bc)}(\psi) 
= \alpha_{21} \kappa^2 \psi^a \bar\psi^b \psi^c \vert w_{3/2} \vert, 
\nonu
\A \A 
\Lambda_5{}^{abc} = \Lambda_5{}^{a(bc)} = \alpha_{22} f_5{}^{a(bc)}(\psi) 
= \alpha_{22} i \kappa^2 \psi^a \bar\psi^b \gamma_5 \psi^c \vert w_{3/2} \vert, 
\nonu
\A \A 
\Lambda^{ab}{}_d{}^{c} = \Lambda^{a[b}{}_d{}^{c]} = \alpha_{23} f^{a[b}{}_d{}^{c]}(\psi) 
= \alpha_{23} i \kappa^2 \psi^a \bar\psi^b \gamma_d \psi^c \vert w_{3/2} \vert, 
\nonu
\A \A 
\Lambda_5{}^{ab}{}_d{}^{c} = \Lambda_5{}^{a(b}{}_d{}^{c)} = \alpha_{24} f_5{}^{a(b}{}_d{}^{c)}(\psi) 
= \alpha_{24} i \kappa^2 \psi^a \bar\psi^b \gamma_5 \gamma_d \psi^c \vert w_{3/2} \vert, 
\nonu
\A \A 
\Lambda^{ab}{}_{de}{}^{c} = \Lambda^{a[b}{}_{de}{}^{c]} = \alpha_{25} f^{a[b}{}_{de}{}^{c]}(\psi) 
= \sqrt{2} \alpha_{25} i \kappa^2 \psi^a \bar\psi^b \sigma_{de} \psi^c \vert w_{3/2} \vert, 
%\nonu
%\A \A 
%\Lambda^{ab}{}_A{}^c = \alpha_{2A} f^{ab}{}_A{}^c(\psi) 
%= \alpha_{2A} \kappa^2 \psi^a \bar\psi^b \gamma_A \psi^c \vert w_{3/2} \vert, 
\nonu
%\A \A 
%C^a{}_A{}^{bc}{}_B{}^d = \alpha_{3AB} b^a{}_A{}^{bc}{}_B{}^d(\psi) 
%= \alpha_{3AB} \kappa^3 \bar\psi^a \gamma_A \psi^b \bar\psi^c \gamma_B \psi^d \vert w_{3/2} \vert, 
%\nonu
\A \A 
\cdots, 
\label{comp1}
\ea
which terminate with bosonic components for the eighth-order functionals of $\psi^a$. 
In these functional components, 
$(\lambda^a, M^{ab}, M_5{}^{ab}, M^a{}_c{}^b, M_5{}^a{}_c{}^b, M^a{}_{cd}{}^b)$ 
are the dynamical vector-spinor and (pseudo) tensor fields 
and $(\Lambda^{abc}, \Lambda_5{}^{abc}, \Lambda^{ab}{}_d{}^{c}, \Lambda_5{}^{ab}{}_d{}^{c}, \Lambda^{ab}{}_{de}{}^{c}, \cdots)$ 
are the general auxiliary fields. 
By using the components in Eq.(\ref{comp1}), the RS fields which constitute the RS action is represented 
by means of the appropriate recombinations of the $\lambda^a$ and the derivatives of the components in $\Lambda_5{}^{ab}{}_d{}^{c}$, 
while the $U(1)$ gauge field is constructed from the components ($M_5{}^a{}_b{}^b$, $M_{5b}{}^{ab}$) as shown below. 

Here we define the following recombinations for the spin-3/2 fields 
by using the components of the fermionic auxiliary fields $\Lambda_5{}^{ab}{}_d{}^{c}$ in Eq.(\ref{comp1}), i.e. 
\ba
\tilde\lambda^a 
\A = \A \lambda^a 
+ {i \over \alpha_{23}} \gamma_5 (a \partial^a \Lambda_5{}^b{}_b{}^c{}_c 
+ b \partial^a \Lambda_5{}^{bc}{}_{bc} 
+ c \partial^a \Lambda_5{}^{bc}{}_{cb} 
+ d \partial_b \Lambda_5{}^c{}_c{}^{ab} 
+ e \partial_b \Lambda_5{}^c{}_c{}^{ba} 
+ f \partial_b \Lambda_5{}^{ca}{}_c{}^b 
\nonu
\A \A 
\hspace{2cm} + g \partial_b \Lambda_5{}^{ac}{}_c{}^b 
+ h \partial_b \Lambda_5{}^{bc}{}_c{}^a 
+ k \partial_b \Lambda_5{}^{acb}{}_c 
+ l \partial_b \Lambda_5{}^{bca}{}_c), 
\label{tilde-lambda0}
\ea
where the coefficients $(a,b,c,\cdots)$ are constrained by means of the $U(1)$ gauge invariance 
for the vector field in the spin-1/2 LSUSY transformations of spin-3/2 fields $\tilde\lambda^a$ in Eq.(\ref{tilde-lambda0}). 
We focus on the variations of the $\tilde\lambda^a$ with respect to (axial) vector components included in $M_5{}^a{}_b{}^c$ in Eq.(22), 
\footnote{
In the limit of spin-1/2 SUSY, the components of $M_5{}^a{}_b{}^c$ include a functional of the spin-1/2 NG fermion 
for a vector field, which has an axial vector nature for $N = 1$ SUSY on the NLSUSY phase \cite{STT}. 
}
which can be estimated by using the following vsLSUSY transformations in Eqs.(\ref{variation-llambda}) and (\ref{variation-lambda}), 
\ba
\A \A 
\delta_\zeta \lambda^a = -{1 \over {4 \alpha_{13}}} \epsilon^{bcde} \partial_b M_5{}^a{}_{fc} \gamma^f \gamma_d \zeta_e 
+ \cdots, 
\label{22}
\\
\A \A 
{1 \over \alpha_{23}} \delta_\zeta \Lambda_5{}^{ab}{}_e{}^c 
= {1 \over \alpha_{13}} M_5{}^b{}_e{}^c \zeta^a 
- {1 \over {4 \alpha_{13}}} (M_5{}^c{}_d{}^a \gamma^d \gamma_e \zeta^b 
+ M_5{}^b{}_d{}^a \gamma^d \gamma_e \zeta^c) 
+ \cdots, 
\label{23}
\ea
%
%%%%%%%%%%%%%%%%%%%%%%%%%%%%%%%%%%%%%%%%
where ellipses mean LSUSY transformation terms with respect to the other dynamical and auxiliary fields 
in the $N = 1$ multiplet (\ref{comp1}) and we use them in LSUSY transformations below 
in order to focus only on the transformation's rule for the spin-(3/2, 1) fields. 

%%%%%%%%%%%%%%%%%%%%%%%%%%%%%%%%%%%%%%%%

In addition, the parts of the spin-1/2 LSUSY transformations in the vsLSUSY ones (\ref{22}) and (\ref{23}) are considered 
by replacing the vsSUSY transformation parameters $\zeta^a$ with $\gamma_5 \gamma^a \zeta$ 
with $\zeta$ being a spin-1/2 SUSY ones, 
though the functionals of the basic component fields in the linearized vsSUSY theory 
remain expressed in terms of the spin-3/2 NG fermions $\psi^a$. 
In fact, the vsLSUSY transformations (\ref{22}) and (\ref{23}) with respect to 
the (axial) vector components $M_5{}^a{}_b{}^b$ and $M_{5b}{}^{ab}$ 
are rewritten by using $\zeta$ as 
\ba
\delta_\zeta \lambda^a 
\A = \A -{i \over {2 \alpha_{13}}} \partial_b M_5{}^a{}_c{}^c \gamma^b \zeta 
- {1 \over {2 \alpha_{13}}} x_0 (\epsilon^{abcd} \partial_b M_{5ce}{}^e - \epsilon^{abcd} \partial_b M_{5ec}{}^e) \gamma_5 \gamma_d \zeta 
\nonu
\A \A 
+ \cdots, 
\label{24}
\\
{1 \over \alpha_{23}} \delta_\zeta \Lambda_5{}^a{}_a{}^b{}_b 
\A = \A {1 \over \alpha_{13}} M_5{}^a{}_b{}^b \gamma_5 \gamma_a \zeta 
- {5 \over {4 \alpha_{13}}} M_{5b}{}^{ab} \gamma_5 \gamma_a \zeta + \cdots, 
\\
{1 \over \alpha_{23}} \delta_\zeta \Lambda_5{}^{ab}{}_{ab} 
\A = \A {1 \over {2 \alpha_{13}}} M_{5b}{}^{ab} \gamma_5 \gamma_a \zeta + \cdots, 
\\
{1 \over \alpha_{23}} \delta_\zeta \Lambda_5{}^{ab}{}_{ba} 
\A = \A {1 \over \alpha_{13}} M_5{}^a{}_b{}^b \gamma_5 \gamma_a \zeta 
- {5 \over {4 \alpha_{13}}} M_{5b}{}^{ab} \gamma_5 \gamma_a \zeta + \cdots, 
\\
{1 \over \alpha_{23}} \delta_\zeta \Lambda_5{}^c{}_c{}^{ab} 
\A = \A -{1 \over {4 \alpha_{13}}} (M_{5c}{}^{ac} \gamma_5 \gamma^b \zeta - M_5{}^b{}_c{}^c \gamma_5 \gamma^a \zeta 
- M_{5c}{}^{bc} \gamma_5 \gamma^a \zeta + \eta^{ab} M_{5c}{}^{dc} \gamma_5 \gamma_d \zeta) 
\nonu
\A \A 
+ {i \over {4 \alpha_{13}}} x_1 \epsilon^{abcd} M_{5ce}{}^e \gamma_d \zeta 
- {i \over {4 \alpha_{13}}} (x_1 + 1) \epsilon^{abcd} M_{5ec}{}^e \gamma_d \zeta + \cdots, 
\\
{1 \over \alpha_{23}} \delta_\zeta \Lambda_5{}^{ca}{}_c{}^b 
\A = \A -{1 \over {4 \alpha_{13}}} (M_5{}^a{}_c{}^c \gamma_5 \gamma^b \zeta + M_5{}^b{}_c{}^c \gamma_5 \gamma^a \zeta) 
\nonu
\A \A 
- {i \over {4 \alpha_{13}}} (x_2 - x_3) (\epsilon^{abcd} M_{5ce}{}^e \gamma_d \zeta 
- \epsilon^{abcd} M_{5ec}{}^e \gamma_d \zeta) + \cdots, 
\\
{1 \over \alpha_{23}} \delta_\zeta \Lambda_5{}^{ac}{}_c{}^b 
\A = \A {1 \over \alpha_{13}} M_5{}^b{}_c{}^c \gamma_5 \gamma^a \zeta 
- {1 \over {4 \alpha_{13}}} M_5{}^a{}_c{}^c \gamma_5 \gamma^b \zeta 
\nonu
\A \A 
+ {i \over {4 \alpha_{13}}} x_4 (\epsilon^{abcd} M_{5ce}{}^e \gamma_d \zeta 
- \epsilon^{abcd} M_{5ec}{}^e \gamma_d \zeta) + \cdots, 
\\
{1 \over \alpha_{23}} \delta_\zeta \Lambda_5{}^{acb}{}_c 
\A = \A {1 \over \alpha_{13}} M_{5c}{}^{bc} \gamma_5 \gamma^a \zeta 
+ {1 \over {2 \alpha_{13}}} M_5{}^a{}_c{}^c \gamma_5 \gamma^b \zeta 
\nonu
\A \A 
- {i \over {2 \alpha_{13}}} x_5 (\epsilon^{abcd} M_{5ce}{}^e \gamma_d \zeta 
- \epsilon^{abcd} M_{5ec}{}^e \gamma_d \zeta) + \cdots. 
\label{31}
\ea
According to these spin-1/2 LSUSY transformations (\ref{24}) to (\ref{31}), 
the variations of $\tilde\lambda^a $ in Eq.(\ref{tilde-lambda0}) are written as 
\ba
\delta_\zeta \tilde\lambda^a 
\A = \A {i \over {4 \alpha_{13}}} \left\{ 4A \ \partial^a M_5{}^b{}_c{}^c 
- (2 - e + B) \partial^b M_5{}^a{}_c{}^c \right. 
\nonu
\A \A 
\hspace{5mm} 
\left. - (5A - 2b + d + e) \partial^a M_{5c}{}^{bc} 
- (d - e - 4l) \partial^b M_{5c}{}^{ac} \right\} \gamma_b \zeta 
\nonu
\A \A 
- {1 \over {4 \alpha_{13}}} \left[ (2x_0 + C x_1 - D) \epsilon^{abcd} \partial_b M_{5ce}{}^e \right. 
\nonu
\A \A 
\hspace{5mm} 
\left. - \{ 2x_0 + C(x_1 + 1) - D \} 
\epsilon^{abcd} \partial_b M_{5ec}{}^e \right] 
\gamma_5 \gamma_d \zeta 
\nonu
\A \A 
+ \cdots, 
\label{LSUSYtilde-lambda0}
\ea
where $A = a + c$, $B = f + g - 4h - 2k$, 
$C = d - e$ and $D = f(x_2 - x_3) - (g - h)x_4 + 2(k - l)x_5$. 

From the variations (\ref{LSUSYtilde-lambda0}) we impose the following conditions 
for the coefficients $(a,b,c,\cdots)$ in the recombinations (\ref{tilde-lambda0}), 
\ba
\A \A 
4A = 2 - e + B, 
\label{c1}
\\
\A \A 
5A - 2b + d + e = -(d - e - 4l), 
\label{c2}
\\
\A \A 
4KA = -(5A - 2b + d + e), 
\label{c3}
\\
\A \A 
K(2 - e + B) = d - e - 4l, 
\label{c4}
\\
\A \A 
K(2x_0 + Cx_1 - D) = -\{ 2x_0 + C(x_1 + 1) - D \}, 
\label{c5}
\ea
so that Eq.(\ref{LSUSYtilde-lambda0}) has a $U(1)$ gauge invariant form. 
Note that a constant $K$ in the conditions (\ref{c1}) to (\ref{c5}) means 
a relative scale between $M_5{}^a{}_b{}^b$ and $M_{5b}{}^{ab}$. 
Also, the condition (\ref{c3}) is equivalent to Eqs.(\ref{c1}), (\ref{c2}) and (\ref{c4}). 
Under those conditions the variations (\ref{LSUSYtilde-lambda0}) become 
\ba
\delta_\zeta \tilde\lambda^a 
\A = \A {i \over \alpha_{13}} A \left\{ \left( \partial^a M_5{}^b{}_c{}^c - \partial^b M_5{}^a{}_c{}^c \right) 
+ K \left( \partial^a M_{5c}{}^{bc} - \partial^b M_{5c}{}^{ac} \right) \right\} \gamma_b \zeta 
\nonu
\A \A 
- {1 \over {4 \alpha_{13}}} (2x_0 + Cx_1 - D) 
\left( \epsilon^{abcd} \partial_b M_{5ce}{}^e + K \epsilon^{abcd} \partial_b M_{5ec}{}^e \right) 
\gamma_5 \gamma_d \zeta + \cdots, 
\ea
which is expressed as 
\be
\delta_\zeta \tilde\lambda^a 
%\A = \A {i \over \alpha_{13}} A \left( \partial^a v^b - \partial^b v^a \right) \gamma_b \zeta 
%- {1 \over {4 \alpha_{13}}} (2x_0 + Cx_1 - D) 
%\epsilon^{abcd} \partial_b v_c \gamma_5 \gamma_d \zeta + \cdots 
%\nonu
%%
= {i \over \alpha_{13}} A F^{ab} \gamma_b \zeta 
- {1 \over {8 \alpha_{13}}} (2x_0 + Cx_1 - D) \epsilon^{abcd} F_{bc} \gamma_5 \gamma_d \zeta + \cdots 
\label{LSUSYtilde-lambda1}
%\nonu
%%
%\A = \A {i \over \alpha_{13}} A F^{ab} \gamma_b \zeta 
%- {1 \over {8 \alpha_{13}}} E \epsilon^{abcd} F_{bc} \gamma_5 \gamma_d \zeta + \cdots, 
\ee
%
%where $E = 2x_0 + Cx_1 - D$. 
%
by defining a vector field $v^a = M_5{}^a{}_b{}^b + K M_{5b}{}^{ab}$ and $F^{ab} = \partial^a v^b - \partial^b v^a$. 

On the other hand, spin-1/2 LSUSY transformations of the vector field $v^a$ are given 
in terms of $\tilde\lambda^a$ as 
\be
\delta_\zeta v^a %\A = \A \delta_\zeta (M_5{}^a{}_b{}^b + K M_{5b}{}^{ab}) 
%\nonu
= -(5 + 2K) \alpha_{13} \bar\zeta \tilde\lambda^a 
+ 2i(1 - 2K) \alpha_{13} \bar\zeta \sigma^a{}_b \tilde\lambda^b + \cdots, 
\label{LSUSYv}
\ee
by replacing the vsSUSY transformation parameters $\zeta^a$ with $\gamma_5 \gamma^a \zeta$ 
in the vsLSUSY transformations (\ref{variation-M0}) for $N = 1$ SUSY. 

Let us use the forms of the LSUSY transformations (\ref{LSUSYtilde-lambda1}) and (\ref{LSUSYv}) 
and estimate the LSUSY invariance of a free action for spin-(3/2, 1) fields, 
\be
S_{(3/2,1)} = \int d^4x \left\{ -{1 \over 4} (F_{ab})^2 
- {1 \over 2} \epsilon^{abcd} \bar{\tilde\lambda}_a \gamma_5 \gamma_b \partial_c \tilde\lambda_d \right\}, 
\label{3/2-1action}
\ee
which is included in the massless limit of the massive spin-3/2 supermultiplets (e.g. \cite{Zi,BGLP}). 
Varying the action (\ref{3/2-1action}) under the LSUSY transformations 
(\ref{LSUSYtilde-lambda1}) and (\ref{LSUSYv}) gives the following terms of $\tilde\lambda^a$ and $v^a$, 
%
%\ba
%
%\delta_\zeta L \A = \A -F^{ab} \partial_a \delta_\zeta v_b 
%- \epsilon^{abcd} \bar\lambda_a \gamma_5 \gamma_b \partial_c \delta_\zeta \lambda_d 
%\nonu
%
%
%\delta_\zeta v^a \A = \A p \bar\zeta \lambda^a + iq \bar\zeta \sigma^a{}_b \lambda^b + \cdots. 
%\nonu
%
%\delta_\zeta \lambda^a \A = \A i r F^{ab} \gamma_b \zeta 
%+ s \epsilon^{abcd} F_{bc} \gamma_5 \gamma_d \zeta + \cdots. 
%\ea
%
%
%%%%%%%%%%%%%%%%%%%%
\ba
\A \A \delta_\zeta S_{(3/2,1)} [{\rm terms\ of\ } \tilde\lambda^a\ {\rm and}\ v^a] 
\nonu
\A \A \hspace{1cm} = \int d^4x \{ -(p + 4s) \bar\zeta \partial_a \tilde\lambda_b F^{ab} 
+ 2i(r - 2s) \bar\zeta \sigma^{ab} \partial_a \tilde\lambda_c F_b{}^c 
\nonu
\A \A 
\hspace{2.7cm} - i(q + 2r - 4s) \bar\zeta \sigma^{ab} \partial_c \tilde\lambda_a F_b{}^c \}, 
\label{variationS}
\ea
%%%%%%%%%%%%%%%%%%%%
where we define 
\ba
p \A = \A - (5 + 2K) \alpha_{13}
\\
q \A = \A 2(1 - 2K) \alpha_{13}
\\
r \A = \A {A \over \alpha_{13}}
\\
s \A = \A - {1 \over {8 \alpha_{13}}} (2x_0 + Cx_1 - D). 
\ea
Therefore, the first three terms in Eq.(\ref{variationS}) vanish at least, provided that 
\ba
\A \A p + 4s = 0, 
\\
\A \A r - 2s = 0, 
\\
\A \A q + 2r - 4s = 0, 
\ea
which mean 
\be
K = {1 \over 2}, \ \ A = 3 \alpha_{13}^2, \ \ 2x_0 + Cx_1 - D = -12 \alpha_{13}^2. 
\label{conditionS}
\ee
Under the conditions (\ref{conditionS}) the LSUSY transformations (\ref{LSUSYtilde-lambda1}) and (\ref{LSUSYv}) become 
\ba
\delta_\zeta \tilde\lambda^a 
\A = \A -{i \over 2} \left( F^{ab} \gamma_b 
- {1 \over 2} \epsilon^{abcd} F_{bc} \gamma_5 \gamma_d \right) \zeta + \cdots, 
\label{LSUSYtilde-lambda2}
\\
\delta_\zeta v^a \A = \A \bar\zeta \tilde\lambda^a + \cdots, 
\label{LSUSYv2}
\ea
where we take the value $\alpha_{13}$ as $\displaystyle{\alpha_{13} = -{1 \over 6}}$. 
Eqs.(\ref{LSUSYtilde-lambda2}) and (\ref{LSUSYv2}) are a well-known form 
as the spin-1/2 LSUSY transformations of the spin-(3/2, 1) fields 
in the massless limit of the massive spin-3/2 supermultiplets. 

%%%%%%%%%%%%%%%%%%%%%%%%%%%%%%%%%%%%%%%%

The arguments for the (vs)LSUSY invariance of a whole action including terms 
with respect to the dynamical and auxiliary fields other than the action (\ref{3/2-1action}) 
under the LSUSY transformations (\ref{LSUSYtilde-lambda2}) and (\ref{LSUSYv2}) including the ellipses, 
which are obtained from the general linear multiplet (\ref{comp}), 
and the relation between (vs)LSUSY actions and the vsNLSUSY one (\ref{vsNLSUSYaction}) are important open problems. 
The spinor superfield formulation of the $N = 1$ gauge (3/2, 1) multiplet \cite{OS} 
may play a crucial role in these problems. 

%%%%%%%%%%%%%%%%%%%%%%%%%%%%%%%%%%%%%%%%

Finally, we mention about the spinorial gauge invariance of the RS action in Eq.(\ref{3/2-1action}) 
together with the $U(1)$ gauge invariance in the linearization of the vsNLSUSY. 
%
%We expect that the spinorial gauge invariance for the spin-3/2 fields $\tilde\lambda^a$ 
%would be realized in the same mechanism as the realization of the $U(1)$ gauge invariance of the action 
%for the vector field $v^a$ in the linearization of the (spin-1/2) VA theory. 
%
In the case of the linearization process of the $d = 2$ VA NLSUSY, e.g. in \cite{MT2,ST6}, 
the minimal (off-shell) $U(1)$ gauge multiplet in the $d = 2$, $N = 2$ LSUSY theory 
is derived from the recombinations of basic fields with much wider number of degrees of freedom, 
which are represented as functionals of the spin-1/2 NG fermions $\psi^i$ ($i = 1,2$). 
Those recombinations can choose the minimal independent part of the general supermultiplet with general auxiliary fields 
so that they play the same role as the WZ gauge removing the redundant degrees of freedom from the wider basic fields. 
The $U(1)$ gauge invariant action which reduces to the VA action in terms of the spin-1/2 goldstinos 
is also constructed in the linearization. 

In addition, the variations of the functional recombinations of $\psi^i$ under the spin-1/2 VA NLSUSY transformations 
with spinor parameteres $\zeta^i$ induce not only the (standard) spin-1/2 LSUSY ones for the $U(1)$ gauge multiplet 
but also the $U(1)$ gauge transformation of the vector field $v^a$ only in the variation of $v^a$ as 
\be
\delta_\zeta v^a(\psi) = -i \epsilon^{ij} \bar\zeta^i \gamma^a \tilde\lambda^j(\psi) + \partial^a W_\zeta (\Lambda^i(\psi)). 
\label{delta-v}
\ee
In Eq.(\ref{delta-v}), $\tilde\lambda^i(\psi)$ are spin-1/2 fields 
defined by means of the following recombinations, 
$\tilde\lambda^i(\psi) = \lambda^i(\psi) + i \!\!\not\!\!\partial \Lambda^i(\psi)$ 
with the auxiliary spinor field $\Lambda^i(\psi)$, 
and $W_\zeta$ means the $U(1)$ gauge transformation parameter 
which is expressed as $W_\zeta = -2 \epsilon^{ij} \bar\zeta^i \Lambda^j(\psi)$. 
In the commutator algebra for $v^a$, 
the cancellation between two gauge transformation terms obtained from the first and second terms of the variation (\ref{delta-v}) occurs 
because of the following relation, 
\be
\delta_{\zeta_1} W_{\zeta_2}(\Lambda^i(\psi)) - \delta_{\zeta_2} W_{\zeta_1}(\Lambda^i(\psi)) 
= -2 (\epsilon^{ij} \bar\zeta_1^i \zeta_2^j A(\psi) + \bar\zeta_1^i \gamma_5 \zeta_2^i \phi(\psi) 
- i \bar\zeta_1^i \gamma^a \zeta_2^i v_a(\psi)) 
\label{commW}
\ee
with ($A$, $\phi$) being scalar and pseudo scalar fields in the $U(1)$ gauge multiplet in $d = 2$. 
%
%Therefore, the commutator algebra does not contain the term for the $U(1)$ gauge transformation 
%in contrast to the case of the WZ gauge; 
%%
%\footnote{
%In the WZ gauge, the commutator algebra for the LSUSY transformation of the vector field 
%induces the $U(1)$ gauge transformation as $[\delta_{\zeta_1}, \delta_{\zeta_2}] = \delta_P + \delta_g$. 
%}
%%
%namely, the form of the commutator algebra for the spin-1/2 NLSUSY transformations, 
%$[\delta_{\zeta_1}, \delta_{\zeta_2}] = \delta_P$, is maintained in the linearization of the VA theory. 
%
The gauge invariant and the redundant sectors are mixed only by means of the gauge transformation 
in the linearization of the VA theory and this fact plays an important role 
to encode the $U(1)$ gauge invariance into the VA theory through the $\delta_\zeta v_a(\psi)$. 

From the above arguments in the linearization of the $d = 2$, $N = 2$ VA NLSUSY theory, 
we anticipate that the same linearization mechanism also works for the vsNLSUSY, 
though the reasonings are merely hypotheses (especially concerning the RS gauge invariance); 
i.e., appropriate recombinations (e.g., in Eq.(\ref{tilde-lambda0})) of the basic fields with general auxiliary fields 
in the linear multiplet (\ref{comp}) would choose an independent minimal off-shell supermultiplet 
including the redefined spin-(3/2, 1) fields ($\tilde\lambda^a(\psi^a)$, $v^a(\psi^a)$) 
in Eqs.(\ref{LSUSYtilde-lambda2}) and (\ref{LSUSYv2}) from the general linear multiplet (\ref{comp}). 

Then, under the vsNLSUSY transformations (\ref{vsNLSUSY}) the variation of $\tilde\lambda^a(\psi^a)$ would induce 
a spinor-gauge transformation term $\partial^a X_\zeta$ in Eq.(\ref{LSUSYtilde-lambda2}) 
with a spinor parameter $X_\zeta$ depending on the auxiliary fields, 
while the variation of $v^a(\psi^a)$ would lead to the same $U(1)$-gauge transformation term $\partial^a W_\zeta$ in Eq.(\ref{LSUSYv2}) 
as in the linearization of the VA NLSUSY theory, 
%
%These gauge transformation terms ($\partial^a X$, $\partial^a W$), 
%which also have the same role to maintain the form of the commutator algebra for the vsSUSY, 
%
%maintain the form of the commutator algebra (\ref{vsNLSUSYcomm}) in the linearization process for vsNLSUSY 
%by means of 
%
which cause cancellations between two (spinor) gauge transformation terms in the commutator algebra for $v^a$ and $\tilde\lambda^a$ 
under the same type relations as Eq.(\ref{commW}). 

%would be included in the ellipses of Eqs.(\ref{LSUSYtilde-lambda2}) and (\ref{LSUSYv2}) 
%and both the spinor and $U(1)$ gauge invariant sectors are connected to the redundant sectors 
%only by means of the gauge transformation parameters ($X$, $W$). 

%At the level of the variations of the functional components of $\psi^a$, 
%the spinor and $U(1)$ gauge invariance, which are not revealed in the vs"NL"SUSY theory, 
%would be built through the terms ($\partial^a X$, $\partial^a W$) 
%in the variations of ($\delta_\zeta \tilde\lambda^a(\psi^a)$, $\delta_\zeta v^a(\psi^a)$) under the vsNLSUSY, 
%though their explicit calculations are future problems. 

%%%%%%%%%%%%%%%%%%%%%%%%%%%%%%%%%%%%%%%%

The results obtained in this letter are summarized as follows. 
We have discussed the linearization of the $N = 1$ and $N$-extended vsNLSUSY based on the commutator algebra (\ref{vsNLSUSYcomm}) 
by focusing on the Baaklini's vsNLSUSY model \cite{Ba}. 
The bosonic and fermionic functionals (\ref{bosonic}) and (\ref{fermionic}) are defined, 
which are composed of products of the spin-3/2 NG fermions $\psi^{(i)a}$ 
and the fundamental determinant $\vert w_{3/2} \vert$ in Eq.(\ref{determinant}). 
By examining the variations and the two supertransformations of the basic component fields (\ref{comp}) 
defined from those functionals under the vsNLSUSY transformations (\ref{vsNLSUSY}), 
we have uniquely determined the general forms of the vsLSUSY transformations 
(\ref{variationD}), (\ref{variation-llambda}), (\ref{variationM}) and (\ref{variation-lambda}), etc. 
The most general structure for the auxiliary fields in Eq.(\ref{comp}) appears in those vsLSUSY transformations. 

In order to study the relation between the linearized vsSUSY theory obtained in this letter 
and supermultiplets with the spin-3/2 fields, 
we have further considered a derivation of spin-1/2 LSUSY transformations of spin-(3/2, 1) fields with U(1) gauge invariance. 
By using the components of the fermionic auxiliary fields $\Lambda_5{}^{ab}{}_d{}^c$ in Eq.(\ref{comp1}), 
the spin-3/2 fields $\tilde\lambda^a$ for $N = 1$ SUSY have been defined in the recombinations (\ref{tilde-lambda0}). 
We have examined the variations of $\tilde\lambda^a$ and shown that they take the $U(1)$ gauge invariant form (\ref{LSUSYtilde-lambda1}) 
under the conditions (\ref{c1}) to (\ref{c5}) for the coefficients $(a,b,c,\cdots)$ in Eq.(\ref{tilde-lambda0}). 

As for the free action (\ref{3/2-1action}) for the spin-(3/2, 1) fields, 
which appears in the massless limit of the massive spin-3/2 supermultiplets, 
we have also discussed the invariance of the action under the LSUSY transformation 
(\ref{LSUSYtilde-lambda1}) and (\ref{LSUSYv}) so that the LSUSY transformations 
(\ref{LSUSYtilde-lambda2}) and (\ref{LSUSYv2}) are obtained under the condition (\ref{conditionS}). 
%%%%%%%%%%%%%%%%%%%%%%%%%%%%%%%%%%%%%%%%%%%%%%%%%%%%%%%%%%%%%%%%%%%%%%
%
The construction of a whole vsLSUSY action including Eq.(\ref{3/2-1action}) 
and the relation between the vsNLSUSY action (\ref{vsNLSUSYaction}) and the vsLSUSY action 
are crucial open problems to be solved in the linearization of vsNLSUSY. 
%
%%%%%%%%%%%%%%%%%%%%%%%%%%%%%%%%%%%%%%%%%%%%%%%%%%%%%%%%%%%%%%%%%%%%%%
The extension of the commutator-based linearization procedure of vsNLSUSY in this letter 
to the vsNLSUSYGR theory with the spin-3/2 NG fermions \cite{ST1} is an interesting problem.

\vspace{7mm}

\noindent
{\large\bf Acknowledgements} \\[2mm]
The author is grateful to the referee for many helpful comments and valuable suggestions.

\newpage

%%%%%%%  References  %%%%%%%%%%%%%%%%%%%%%%%%%%%%%%%%%%%%%%%
%
\newcommand{\NP}[1]{{\it Nucl.\ Phys.\ }{\bf #1}}
\newcommand{\PL}[1]{{\it Phys.\ Lett.\ }{\bf #1}}
\newcommand{\CMP}[1]{{\it Commun.\ Math.\ Phys.\ }{\bf #1}}
\newcommand{\MPL}[1]{{\it Mod.\ Phys.\ Lett.\ }{\bf #1}}
\newcommand{\IJMP}[1]{{\it Int.\ J. Mod.\ Phys.\ }{\bf #1}}
\newcommand{\PR}[1]{{\it Phys.\ Rev.\ }{\bf #1}}
\newcommand{\PRL}[1]{{\it Phys.\ Rev.\ Lett.\ }{\bf #1}}
\newcommand{\PTP}[1]{{\it Prog.\ Theor.\ Phys.\ }{\bf #1}}
\newcommand{\PTPS}[1]{{\it Prog.\ Theor.\ Phys.\ Suppl.\ }{\bf #1}}
\newcommand{\AP}[1]{{\it Ann.\ Phys.\ }{\bf #1}}

\end{document}